\documentclass[referee]{aa} % for a referee version
\input psfig.sty
\begin{document}
\title{Radio Spectral Index Study of the SNRs OA184 and VRO42.05.01}
\author{Denis Leahy\inst{1}  
 \and
        Wenwu Tian\inst{1,2}}
\authorrunning{D. Leahy and W.W. Tian} 
\offprints{D. Leahy and W. W. Tian}
\institute{Department of Physics \& Astronomy, University of Calgary, Calgary, Alberta T2N 1N4, Canada\\
\and 
 National Astronomical Observatories, CAS, Beijing 100012, China}

\date{Received XX, 2005; accepted XX, 2005} 
\abstract{
New images of the Supernova Remnants (SNRs) OA184 and 
VRO42.05.01 are presented at 408 MHz and 1420 MHz, 
from the Canadian Galactic Plane Survey (CGPS) data. 
The SNRs' flux densities at both 408 MHz and 1420 MHz are found and
corrected for flux densities from compact sources within the SNRs. 
The integrated flux density based spectral indices (S$_{\nu}$$\propto$$\nu$$^{-\alpha}$)  
are 0.25$\pm$0.03 for OA184 and 0.36$\pm$0.06 for VRO42.05.01. 
These agree with the respective T-T plot spectral indexes 
of 0.23$\pm$0.06 and 0.36$\pm$0.03. 
OA184's spectral index is smaller than previously published values. 
The older flux density values of OA184 from lower resolution data 
include contributions from a non-SNR ridge emission region and from
compact sources within OA184.  
Subtracting these contributions
results in a spectral index of 0.32$\pm0.06$ (38 MHz to 2695 MHz) 
or 0.28$\pm0.06$ (408 MHz to 2695 MHz).
Correction of published flux densities for compact sources for VRO.42.05.01 
results in a spectral index of 0.32$\pm0.05$ for 38 MHz to 2695 MHz.
We also find spatial variations of spectral index. For OA184 $\alpha$ varies 
from 0.1 to 0.3 (with errors $\simeq$0.1).  
For VRO42.05.01, the shell region has $\alpha$=0.31 and the wing region has 
$\alpha=0.47$ (with errors $\simeq$0.03). 
}
\titlerunning{Spectrum of the SNRs OA184 and VRO42.04.01}
\maketitle 

\section{Introduction}
The paper is part of a continuing study of supernova remnants' %(SNRs) 
spectral index variation. 
The radio spectra of the SNRs OA184 and VRO42.05.01 have not been 
studied before in detail.  
However, both SNRs' basic physical features have been studied in  
previous research. 

 The distance to OA184 was first suggested to be 8$\pm$2 kpc (Routledge et al. 1986)
based on the surface brightness-diameter ($\Sigma$-D) relation, then argued to be 
about 2.6 kpc by Leahy and Marshall (1988), based on infrared observations. 
The most reliable estimate is
4.5$\pm$1.5 kpc by Landecker et al. (1989) based on HI measurements. 
The latter yields a radius for OA184 of $\simeq$50pc.
OA184 is roughly elliptical in shape in 
both radio and optical images with similar structure (Routledge et al. 1986) 
and has no molecular clouds nearby (Huang and Thaddeus 1986). It is just 
past the end of the adiabatic phase of evolution (Leahy and Marshall 1988). 

VRO42.05.01 has an unusual shape with a northeastern shell intersected 
by a much larger bowl-shaped wing in the southwest (Landecker et al. 1982).
 The shell has a radius of 25 pc, for a distance of 5 kpc, and the SNR is 
currently in its isothermal stage (Pineault et al. 1987).
The SNR is interacting with an HI cloud and is inside an expanding HI shell, which 
yields a more reliable distance of 4.5$\pm$1.5 kpc (Landecker et al. 1989).
Burrows and Guo (1994) present the first X-ray images of VRO.42.05.01: in X-rays 
it is center-filled. The X-ray flux and temperature yield an improved age: 13000 
to 24000 yr for models without and with clouds.

In this paper, OA184 and VRO42.05.01 are mapped at higher 
sensitivity than previously at 408 MHz and 1420 MHz.  The resolution is also 
higher than that for any previous map, except for the 1420 MHz map of 
Pineault et al. (1987) for  VRO42.05.01.  The spectral indexes are determined,  
including an analysis with archival datasets.
    
\section{Observations and Image Analysis}

The 408 MHz and 1420 MHz data sets come from the CGPS,
which is described in detail by Taylor et al. (2003).
The data sets are mainly based on observations from the Synthesis Telescope 
(ST) of the Dominion Radio Astrophysical Observatory (DRAO). The spatial
resolution is better than 1'$\times$ 1' cosec($\delta$) 
at 1420 MHz and 3.4'$\times$3.4' cosec($\delta$) at 408 MHz.  
DRAO ST observations 
are not sensitive to structures larger than an angular 
size scale of about 3.3$^{o}$ at 408 MH and 56' at 1420 MHz. Thus the CGPS includes 
data from the 408 MHz all-sky survey of Haslam et al (1982) which has an 
effective resolution of 51' and the Effelsberg 1.4 GHz Galactic plane survey 
of Reich et al. (1990, 1997) with resolution 9.4' for large scale emission 
(the single-dish data are freely available by http://www.mpifr-bonn.mpg.de/survey.html). 

We check the OA184 and VRO40.05.01 maps from the 408 MHz all-sky survey, and 
find that there exists an artifact, i.e. low-level striping, that appears as 
discontinuities across lines of constant Right Ascension. The maximum 
amplitude of the striping, with a scale of 3$^{0}$, is less than 5$\%$. 
In the 408 MHz single-dish map, OA184 is located between the lowest and 
maximum amplitude of one strip, and is weakly contaminated from 0 in the high RA 
side to about 5$\%$ in the low RA side by the artifact.  
VRO40.05.01 is located around the top of one strip, and is contaminated from 
3 to 5$\%$ by the artifact. In the 1420 MHz Effelsberg map of the both SNRs, 
there do not appear to be any significant sidelobe effects. 

Estimating the SNRs' flux densities at each frequency is challenging and important 
for the study of the radio spectrum. We analyze the images and determine flux
densities using the DRAO export software package.  
 For the SNRs, integrated flux densities are calculated using the madr routine and
errors are found by comparing results for several different choices of background
region. For compact sources, the flux densities are calculated using the
fluxfit routine, and errors are taken as the formal Gaussian fit errors. 
The influence of compact sources within the SNRs is very much 
reduced by employing similar methods to Tian and Leahy (2005).
 
\section{Results}
\begin{figure*}
\vspace{90mm}
\begin{picture}(200,300)
\put(-30,215){\includegraphics{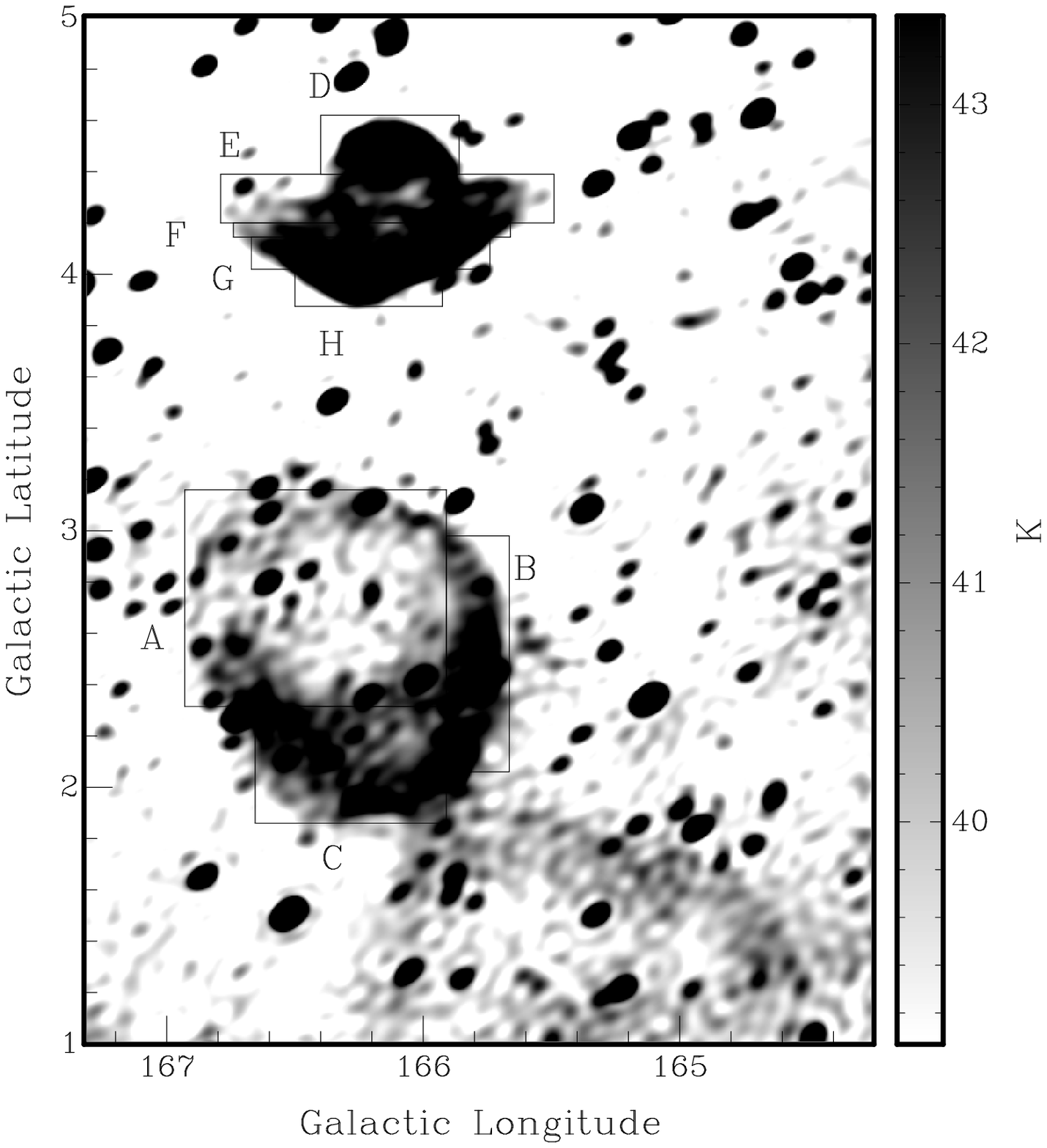}}
\put(220,215){\includegraphics{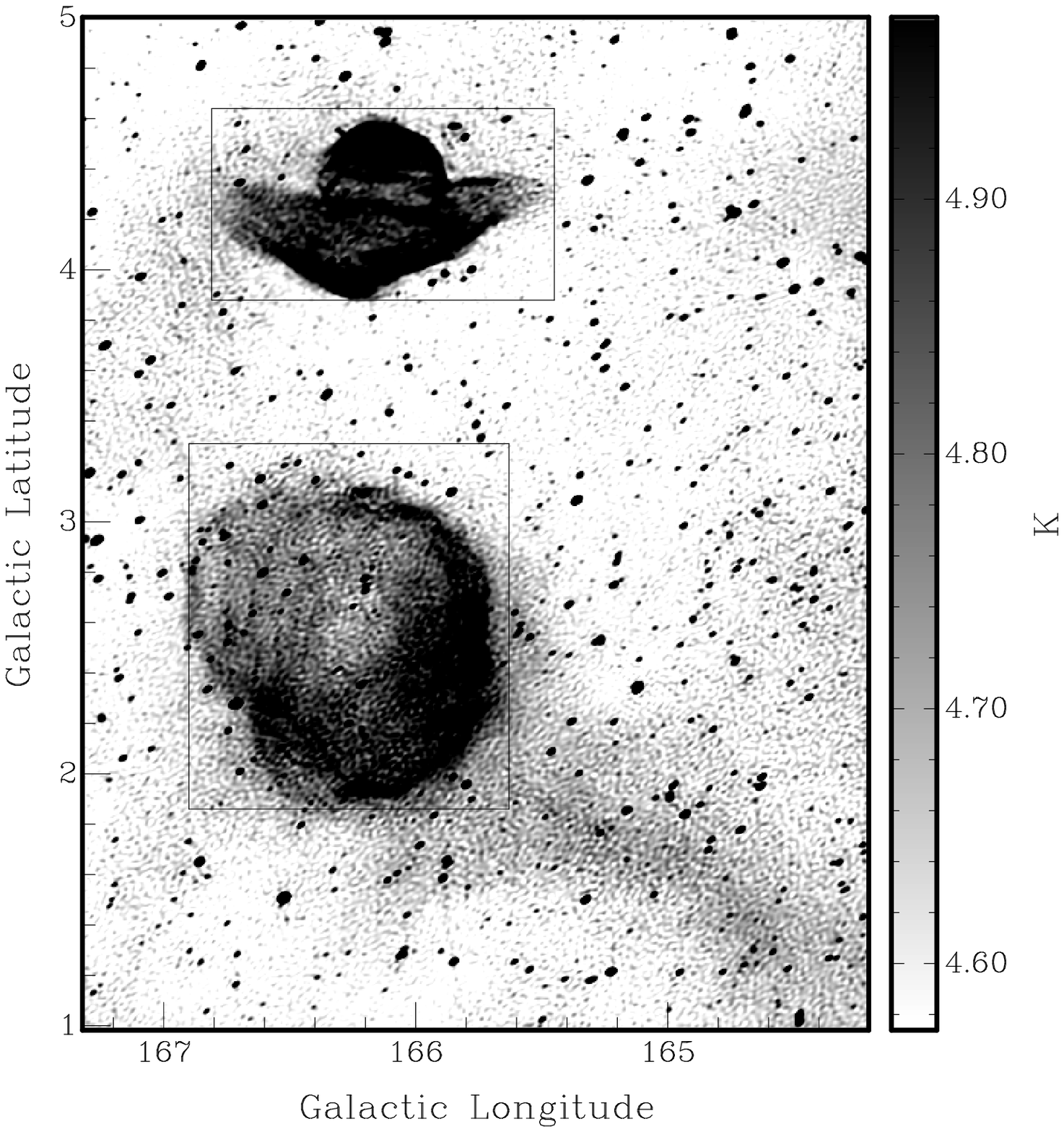}}
\put(-50,-60){\includegraphics{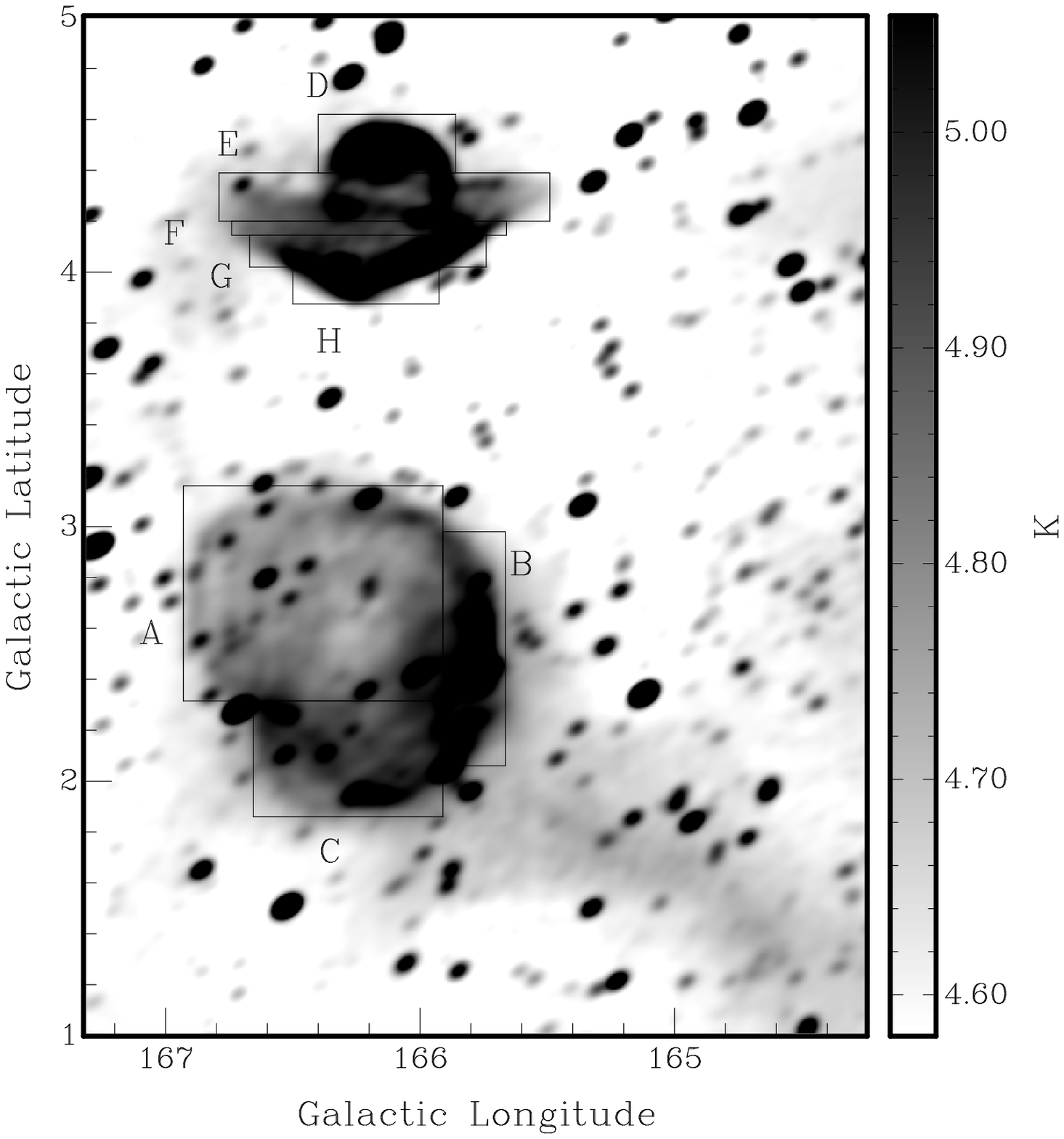}}
\put(196,-82){\includegraphics{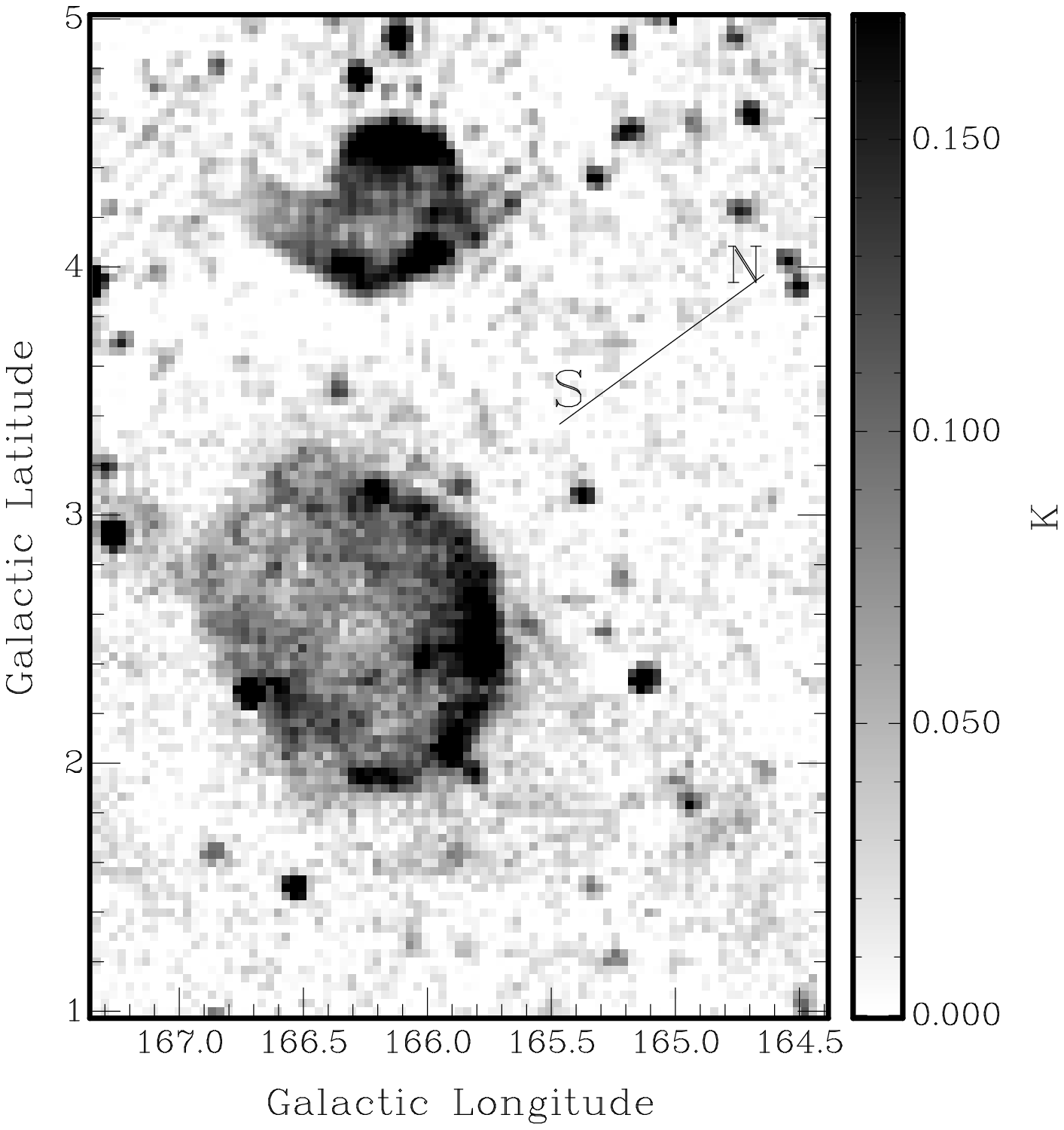}}
\end{picture}
\caption[xx]{In each image OA184 is in the lower half and VRO42.05.01 
is in the upper half. The first row of images shows the CGPS maps at 
408 MHz (left) and 1420 MHz (right). The second row shows the 1420 MHz 
map convolved to the same resolution as the 408 MHz map (left) and the 
2695 MHz Effelsberg map (right). The two boxes used for whole SNR T-T plots 
are shown in the upper right image. The 8 boxes, labeled with letters and 
used for SNR sub-areas T-T plots, are shown in the upper and lower left images.
The direction of North (N) and South (S) is marked on the lower right image.}
\end{figure*}

\subsection{Structure at 408 MHz and 1420 MHz}
The CGPS images at 408 MHz and 1420 MHz are shown in the upper left and right
panels of Fig. 1. OA184 is in the lower half and VRO42.05.01 
is in the upper half of each image.
The lower left panel shows the 1420 MHz map which has been convolved to the same
resolution as the 408 MHz map prior to source removal. For reference we also
reproduce the 2695 MHz Effelsberg map (lower right; F\"urst et al., 1990) in Fig. 1. 
The Effelsberg map has an resolution of 4.3' and a sensitivity of 50 mKT$_{B}$.

The high resolution 408 MHz image for OA184 is the first published at that frequency. 
Its large elliptical shell-type outline is clearly seen. Due to the high 
sensitivity and dynamic range of these data, weak diffuse emission of the eastern 
part of this low brightness SNR is detected. 
Compact sources distributed across the face of the SNR area are prominent. 
One large diffuse emission feature just west of OA184 is detected, which 
has not been reported previously.   
The 1420 MHz map (expanded view shown in Fig. 2)
shows both weak eastern and bright western SNR 
features much better than previous observations.
The features of OA184 at 408 MHz and 1420 MHz are very similar.
The weak diffuse region west of OA184 is also seen clearly at 1420 MHz, with more detail 
at 1420 MHz than at 408 MHz. 
This confirms that the new ridge feature is real. 

\begin{figure*}
\vspace{55mm}
\begin{picture}(50,50)
\put(-30,-100){\includegraphics{oa184-1420.eps}}
\put(240,-93){\includegraphics{vor_1420.eps}}
\end{picture}
\caption[xx]{Expanded 1420 MHz images of OA184 (the left) and VRO42.05.01 (the right)  with larger greyscale range.}
\end{figure*}

VRO42.05.01 is significantly brighter than OA184.
Fig. 1 shows the fainter features at 408 and 1420 MHz.
Fig. 2 (right) gives the 1420 MHz higher resolution image with larger range of brightness temperature
so that more detail can be seen in the bright regions.
The 1420 MHz image best reveals its unusual shape: a northeastern 
hemispherical `shell' joined to a much larger southwestern bowl-shaped 
`wing'. 
The brightest parts of the SNR are at the edge of the shell and at the edge of the wing.
There is a prominent bridge, not previously noted, along the bottom edge of the shell,
seen clearly in Fig. 1 just below the faintest part of the shell interior. 
The 1420 MHz map in Fig. 2 reveals 
more diffuse emission than previously seen in both the southeastern and northwestern ends of the 
wing due to the higher dynamic range of the current data. At 1420 MHz
detailed filamentary structure is seen over much of the shell and in the bottom part of
the wing region. The 1420 MHz
map clearly shows that the boundary between the wing and the shell is 
aligned parallel to the galactic plane, as described first by Landecker et al. (1982). 
The radio structures of VRO42.05.01 are the same at 408 MHz and 
1420 MHz up to the limit of the resolution of the 408 MHz map.   

\begin{table*}
\begin{center}
\caption{List of brightest compact sources and their Flux Densities (FD) inside OA184 and VRO42.05.01}
\setlength{\tabcolsep}{1mm}
\begin{tabular}{cccccc}
\hline
\hline
 Source Num. &GLONG& GLAT & FD at 408MHz& FD at 1420MHz& Sp. Index\\
\hline
 &deg &deg &mJy&mJy&$\alpha$\\
\hline
\hline
For OA184\\
\hline
1&   165.887&2.055 &664 $\pm$20&229$\pm$ 9&0.86 ( 0.81 to 0.90)\\
2&   165.735&2.450 &309 $\pm$33&273$\pm$22&0.10 (-0.05 to 0.25)\\
3&   166.198&3.120 &196 $\pm$ 8& 93$\pm$ 4&0.60 ( 0.56 to 0.65)\\
4&   166.006&2.425 &182 $\pm$ 6& 88$\pm$ 3&0.58 ( 0.54 to 0.62)\\
5&   166.605&2.803 &150 $\pm$10& 60$\pm$ 3&0.74 ( 0.67 to 0.81)\\
6&   166.615&3.175 &121 $\pm$ 7& 53$\pm$ 3&0.66 ( 0.60 to 0.73)\\
7&   166.604&3.077 & 96 $\pm$ 7& 24$\pm$ 3&1.10 ( 0.99 to 1.23)\\
8&   166.211&2.359 & 90 $\pm$ 6& 36$\pm$ 2&1.74 ( 1.68 to 0.81)\\
9&  166.195&2.768 & 59 $\pm$ 4& 31$\pm$ 1&0.67 ( 0.61 to  0.75)\\
10&  166.440&2.850 & 50 $\pm$ 4& 28$\pm$ 1&0.46 ( 0.40 to  0.52)\\
11&  165.990&3.036 & 47 $\pm$14& 14$\pm$ 5&1.00 ( 0.68 to  1.53)\\
12&  166.391&3.169 & 45 $\pm$ 5& 17$\pm$ 2&0.78 ( 0.66 to  0.92)\\
13&  166.535&2.102 & 45 $\pm$ 3& 31$\pm$ 1&0.30 ( 0.25 to  0.36)\\
14&  166.821&2.345 & 40 $\pm$ 5& 26$\pm$ 1&0.35 ( 0.25 to  0.47)\\
15&  166.878&2.826 & 39 $\pm$10& 20$\pm$ 4&0.54 ( 0.32 to 0.85)\\
16&  165.895&2.316 & 26 $\pm$ 7& 14$\pm$ 4&0.49 ( 0.24 to  0.84)\\
17&  165.926&1.839 & 20 $\pm$ 2& 10$\pm$ 1&0.60 ( 0.50 to 0.71)\\
\hline
For VRO42.05.01 \\
\hline
1&   166.327&4.024 &151 $\pm$17& 70$\pm$10&0.62 ( 0.41 to 0.83)\\
2&  166.327&4.040 & 73 $\pm$16& 35$\pm$10&0.59 ( 0.34 to 0.96)\\
3&  166.696&4.351 & 53 $\pm$ 4& 26$\pm$ 1&0.57 ( 0.50 to 0.65)\\
\hline
\hline
\end{tabular}
\end{center}
\end{table*}

\begin{table}
\begin{center}
\caption{408-1420 MHz T-T Plot spectral indices
with and without Compact Sources(CS)}
\setlength{\tabcolsep}{1mm}
\begin{tabular}{cccc}
\hline
\hline
 Sp. Index   |    &$\alpha$ & $\alpha$& $\alpha$ \\
\hline
\hline
 Area \vline &including CS & CS subtracted & CS removed \\
\hline
 A& 0.43$\pm$0.16& 0.21$\pm$0.26&0.22$\pm$0.29\\
 B& 0.19$\pm$0.04& 0.19$\pm$0.10&0.11$\pm$0.05\\
 C& 0.45$\pm$0.09& 0.31$\pm$0.28&0.29$\pm$0.16\\
 \hline
 All OA184& 0.28$\pm$0.10& 0.21$\pm$0.12 & 0.23$\pm$0.06\\
\hline
 D& 0.31$\pm$0.01& 0.31$\pm$0.01&0.31$\pm$0.01\\
 E& 0.35$\pm$0.06& 0.39$\pm$0.11&0.39$\pm$0.08\\
 F& 0.53$\pm$0.04& 0.53$\pm$0.04&0.53$\pm$0.04\\
 G& 0.49$\pm$0.02& 0.44$\pm$0.01&0.45$\pm$0.01\\
 H& 0.48$\pm$0.01& 0.47$\pm$0.01&0.47$\pm$0.01\\
\hline
All VRO42.05.01&0.38$\pm$0.03&0.36$\pm$0.03&0.36$\pm$0.03\\
\hline
 \hline
\end{tabular}
\end{center}
\end{table}

\begin{table*}
\begin{center}
\caption{Integrated flux densities and spectral indices of OA184, VRO42.05.01, and 
compact sources within OA184 and VRO42.05.01}
\setlength{\tabcolsep}{1mm}
\begin{tabular}{ccccccc}
\hline
\hline
Freq.& OA184& CS of OA184&OA184+CS& VRO42 & CS of VRO42 & VRO42 +CS\\
\hline
MHz &Jy&Jy &Jy&Jy&Jy&Jy\\
\hline
\hline
 408& 10.7$\pm$1.0&2.2$\pm$0.2&12.9$\pm$1.2&8.1$\pm$1.0 &0.28$\pm$0.04&8.4$\pm$1.0\\
 1420&7.8$\pm$0.3 &1.0$\pm$0.1&8.8$\pm$0.4 &5.2$\pm$0.2 &0.13$\pm$0.02&5.3$\pm$0.2\\
\hline
$\alpha$&0.25$\pm$0.03&0.63$\pm$0.12&0.31$\pm$0.04&0.36$\pm$0.06 &0.62$\pm$0.12 &0.37$\pm$0.05 \\
\hline
\hline
\end{tabular}
\end{center}
\end{table*}

\begin{table}
\begin{center}
\caption{Integrated Flux Densities (FD) of OA184}
\setlength{\tabcolsep}{1mm}
\begin{tabular}{ccccc}
\hline
\hline
Freq. &Beamwidth&FD & references\\
MHz   &arcmin   & Jy               & \\
\hline
\hline
  38.0 & 45$\times$45 & 70.0* $\pm$20.0& 1971, Haslam and Salter\\
 102.0 & 48$\times$25 & 37.0 $\pm$6.0 & 1994, Kovalenko et al.\\
 408.0 &3.4$\times$5.1& 10.7 $\pm$1.0 & this paper\\
 408.0 & 45$\times$45 & 15.5 $\pm$4.0 & 1971, Haslam and Salter \\
 610.0 & 15$\times$15 & 13.5 $\pm$ 1.5 & 1965, Dickel and Yang\\
 1420.0& 1$\times$1.4 & 9.0  $\pm$ 0.5 & 1986, Routledge et al.\\
 1420.0& 1$\times$1.5 & 7.8  $\pm$ 0.3 & this paper\\
 2695 & 3.4$\times$3.4& 6.9  $\pm$ 0.7 & see text \\
 2700.0& 5$\times$5   & 6.0  $\pm$ 2.2 & 1973, Willis \\
\hline
\hline
& & &*corrected using scale correction \\
& & &of Roger et al. 1973 \\
\end{tabular}

\caption{Integrated flux densities (FD) of VRO42.05.01}
\setlength{\tabcolsep}{1mm}
\begin{tabular}{ccccc}
\hline
\hline
Freq. &Beamwidth &FD & references\\
MHz   &arcmin    & Jy               & \\
\hline
\hline
  38.0 & 45$\times$45 & 22.4*        & 1966, Williams et al.\\
  83.0 & 59$\times$31 & 20.0 $\pm$7.0& 1994, Kovalenko et al.\\
 408.0 & 45$\times$45 & 9.0 $\pm$1.6 & 1971, Haslam and Salter\\
 408.0 &3.4$\times$5.1& 8.1 $\pm$1.0 & this paper\\
 610.0 & 15$\times$15 & 7.5 $\pm$1.0 & 1965, Dickel et al.\\
 1400.0& 10$\times$10 & 6.8          & 1972, Felli and Churchwell\\
 1420.0& 1$\times$1.4 & 6.0 $\pm$0.8 & 1982, Landecker et al.\\
 1420.0& 1$\times$1.5 & 5.2 $\pm$0.2 & this paper\\
 2965& 3.4$\times$3.4 & 4.6 $\pm$0.2 & see text \\
 2700.0& 5$\times$5   & 5.2** $\pm$1.0 & 1973, Willis \\
\hline
\hline
& & &*corrected using scale correction \\
& & &of Roger et al. 1973. \\
& & &**probably high due to inclusion\\
& & &of compact sources to the northeast.\\
\end{tabular}
\end{center}
\end{table}

\subsection{T-T plot Spectral Indices}
Bright compact sources effect both integrated flux density for the SNRs and 
spectral indices, we correct for the effects of compact sources below. 
 Each compact source is fit by 1 to 3 Gaussian components plus twisted plane
background (details can be found in the DRAO export package manual). Each 
Gaussian has 6 parameters: x and y positions, normalization, major and minor axes,
and orientation of major axis, resulting in up to 18 free parameters fit to each
compact source.
Table 1 lists properties of the 20 brightest compact sources which are detected
within OA184 and VRO42.05.01 at both frequencies.   

First we discuss spectral indices between 408 MHz and 1420 MHz based on 
the T-T plot method.
The principle of the T-T plot method is that spectral indices 
(T$_{\nu}$=T$_{o}$$\nu$$^{-\beta}$) are calculated from a fit of a linear 
relation to the T$_{1}$-T$_{2}$ values of all pixels within a given map region. 
T$_{1}$ is the brightness temperature of a map pixel at one frequency and 
T$_{2}$ is for the second frequency. The brightness temperature spectral 
index $\beta$ is derived from the slope of the curve. 
The flux density spectral index $\alpha$ 
(S$_{\nu}$$\propto$$\nu$$^{-\alpha}$) 
is related to $\beta$ by $\beta$=$\alpha$+2. Spectral index 
refers to flux density spectral index $\alpha$ in this paper unless specifically 
noted otherwise. The T-T plot method has been widely used for spectral index 
calculation (e.g., Zhang et al., 1997; Leahy and Roger, 1998). 

For the T-T plot analysis, first a single region for the whole of each SNR is used, 
as shown in Fig. 1. These two regions yield the T-T plots shown in Fig. 3: for OA184 
in the top row and for VRO42.05.01 in the bottom row. 
Three cases are considered: using all pixels including compact sources; using all 
pixels after subtracting Gaussian fits to the compact sources 
listed in Table 1 from the images; 
and excluding compact sources. 
For OA184 the compact sources are bright compared to the SNR emission. 
Since the compact sources have a steeper spectrum than the SNR, 
they are seen in the T-T plot (Fig. 3 upper left panel) as the steeper lines of 
points extending to higher $T_B$. Subtracting compact sources from the image before making
the T-T plot removes the lines of points associated with the compact sources
(Fig. 3 upper-middle panel). However the Gaussian subtraction is imperfect, leaving 
artifacts in the T-T plot. 
There are two reasons why the subtraction is imperfect. 
First, the maps are made using CLEAN, and bright compact (but not necessarily point) 
sources can have up to 
$\approx$100 CLEAN components, so that with a few (1-3) Gaussians does not 
perfectly reproduce 
the source. Second, the Gaussian fitting routine does not work perfectly in the 
presence of structured diffuse emission. 
The next step is to completely remove regions of pixels including 
the compact sources from the analysis. Each region is taken to be a few 
beamwidths across, so that any contribution from the compact source is below 
1$\%$ of the diffuse SNR emission. Thus any artifacts associated with the compact
source are also removed. 
The last step produces the safest results (upper rightmost plot of Fig. 3).
The effects are not as visible for VRO42.05.01 (lower plots of Fig. 3) 
since the compact source flux density
is much weaker relative to the SNR emission compared to OA184.  

\begin{figure*}
\vspace{100mm}
\begin{picture}(60,100)
\put(-98,405){\includegraphics{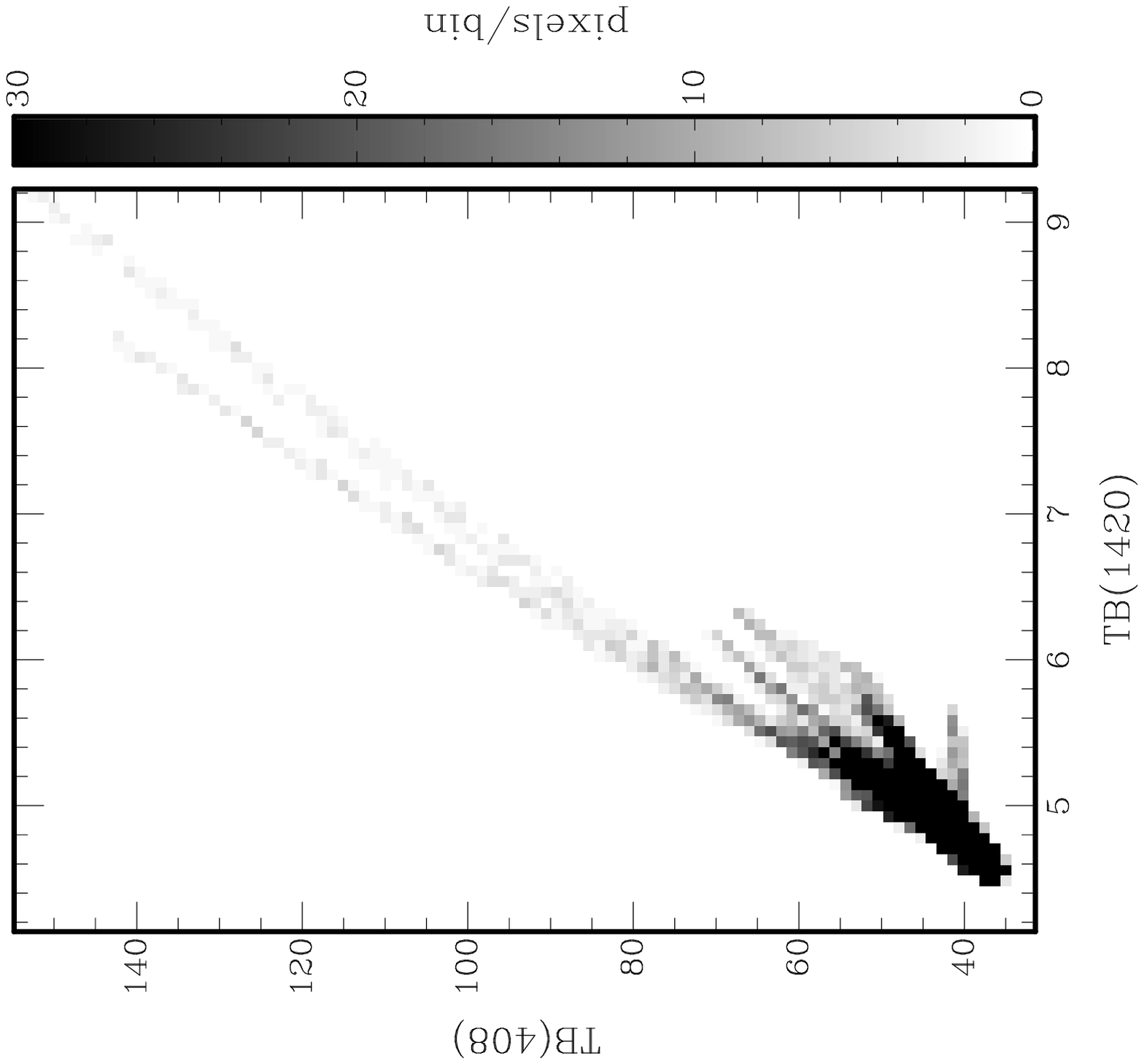}} 
\put(72,405){\includegraphics{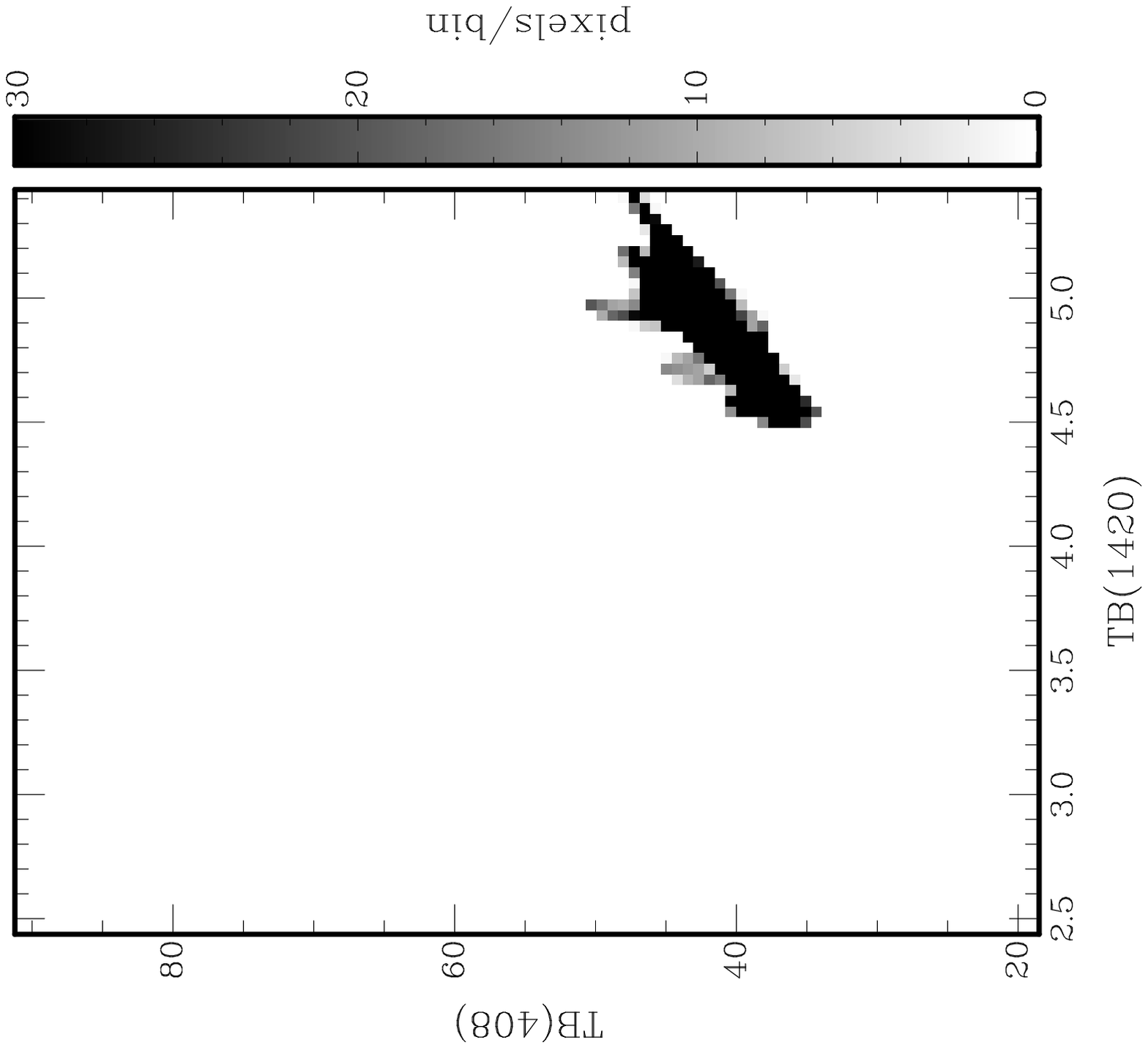}} 
\put(248,405){\includegraphics{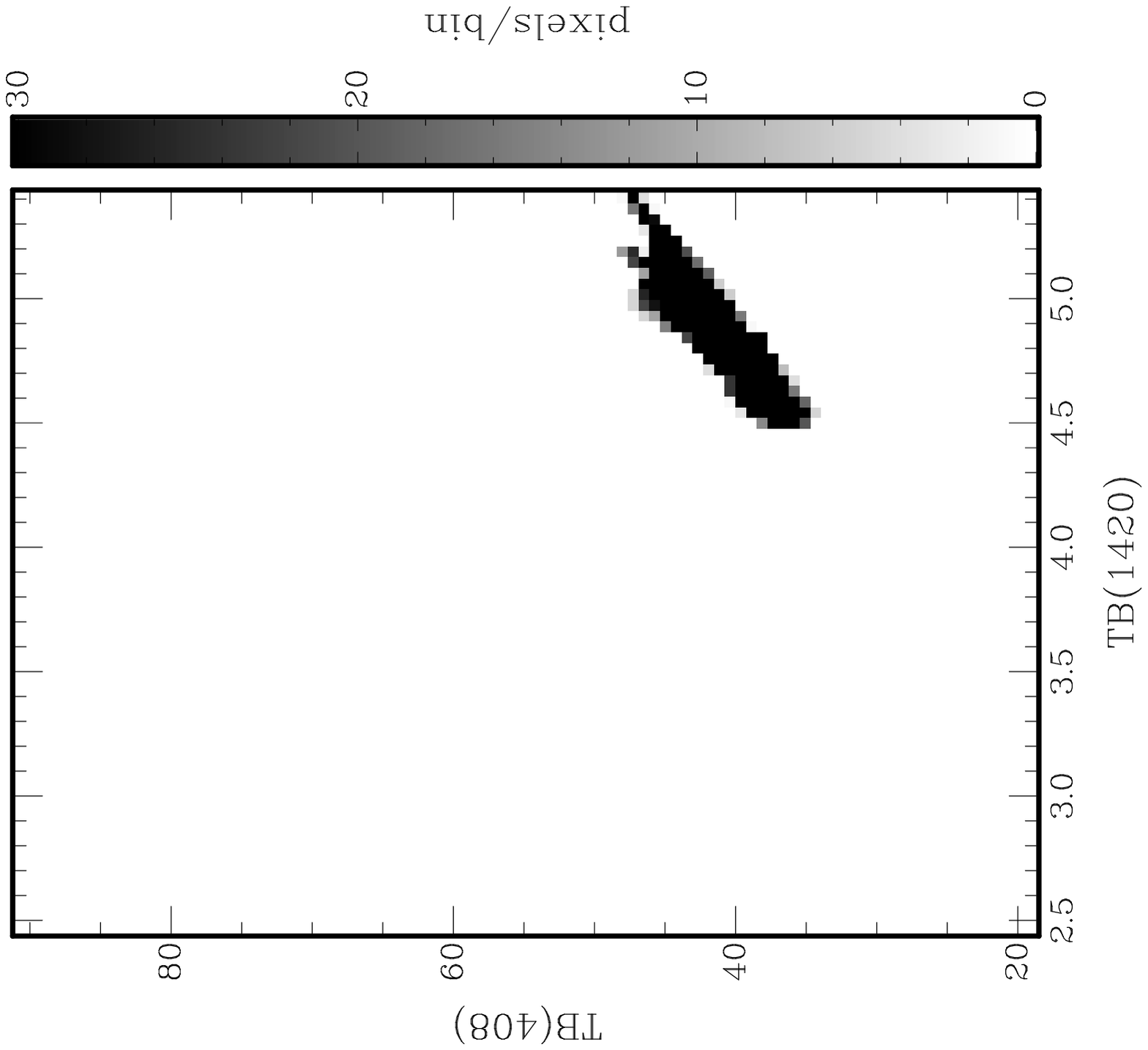}}
\put(-65,190){\includegraphics{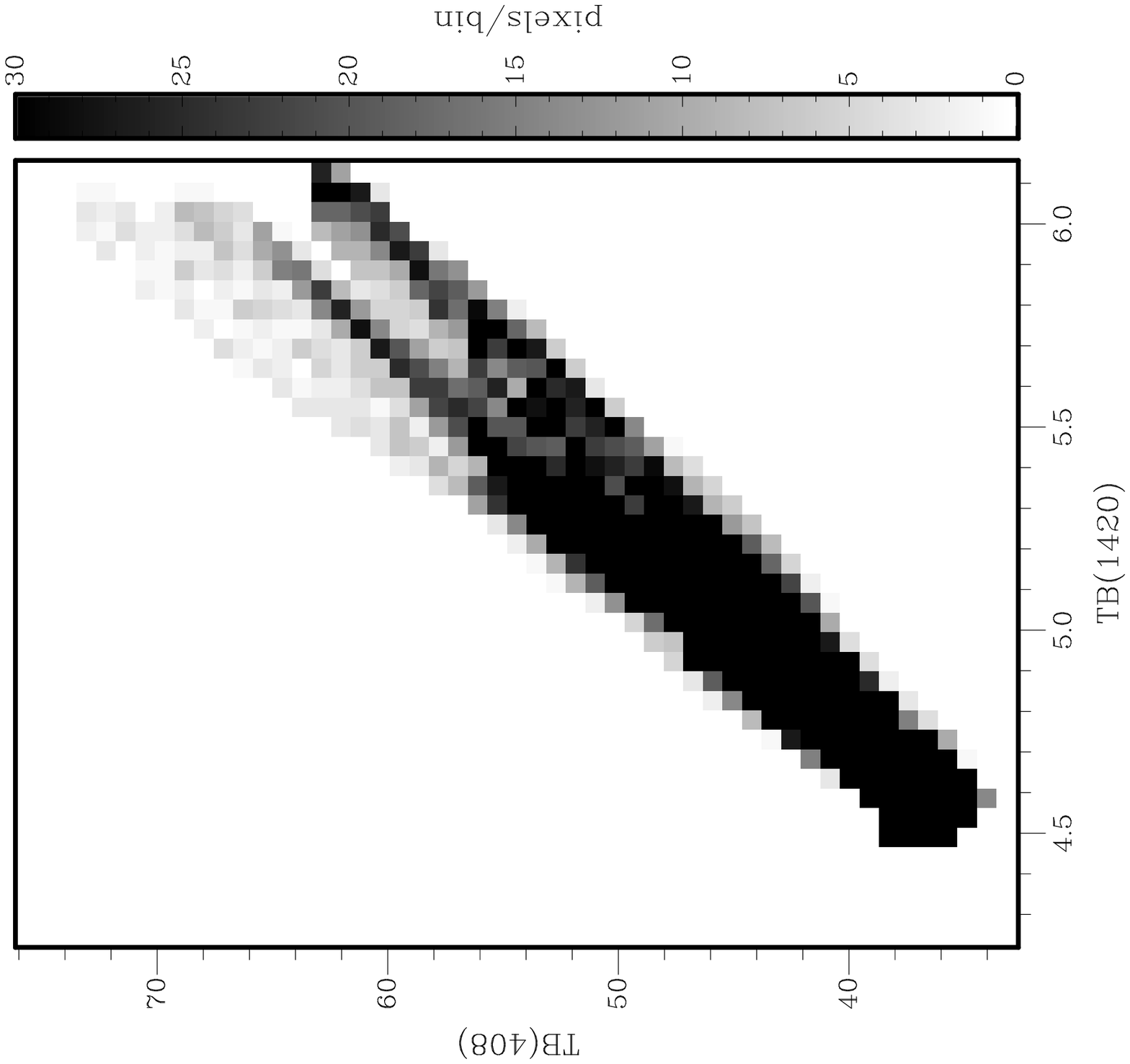}}
\put(108,190){\includegraphics{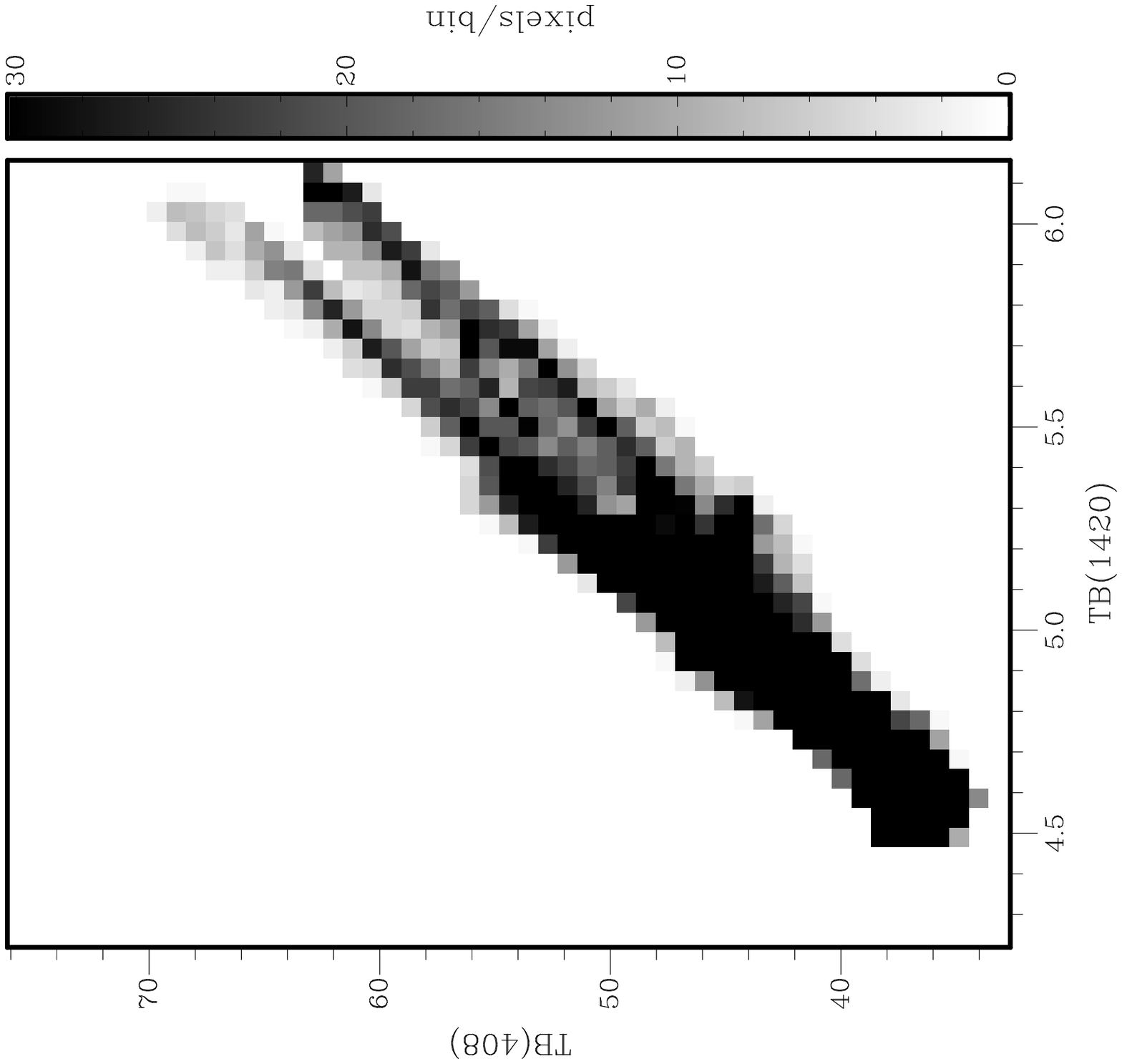}}
\put(280,190){\includegraphics{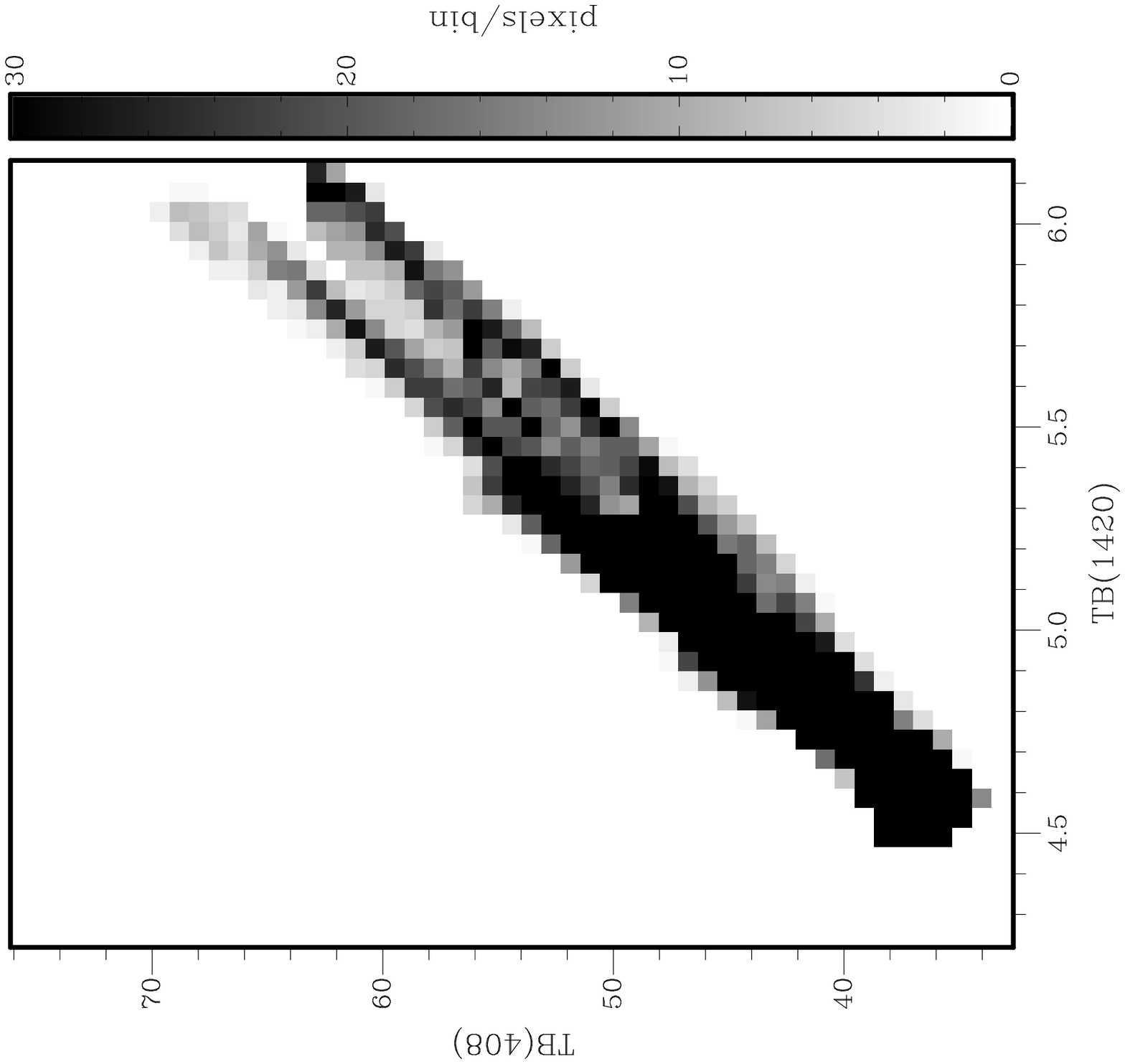}}
\end{picture}
\caption[xx]{ Whole SNR 408-1420 MHz T-T plots for OA184 (first row) and 
VRO42.05.01 (second row).
From left to right for OA184: plot for map including compact sources 
($\alpha$=0.28 $\pm$0.10); 
plot for map with Gaussian fits to compact sources subtracted ($\alpha$=0.21$\pm$0.12); 
plot for compact sources removed from analysis (an averaged $\alpha$ is 0.23$\pm$0.06). The respective values of $\alpha$ for VRO42.05.01 are 0.38$\pm$0.03, 0.36$\pm$0.03 and 0.36$\pm$0.03.}
\end{figure*}

Next, the SNRs are subdivided into smaller areas, 
labeled A to H in the left panels of
Fig. 1, to search for spatial variations in spectral index.  
Table 2 lists the results for three cases of analysis: 
including compact sources, subtracting compact sources, and removing compact 
sources. There is not a large difference in results between the compact sources
subtracted and compact sources removed methods. 
 The largest difference is for area B: here there is a 0.7$\sigma$ difference
between the two methods.
Visual inspection of the T-T plots shows that the third method produces the most reliable results, as it did for the entire SNR. 
The compact sources' influence on the spectral index calculation 
is obvious in the T-T plots. From now on we discuss
spectral indices derived with compact sources removed, unless specified otherwise.   

\subsection{Integrated Flux Densities and Spectral Indices}
We have derived integrated flux densities for OA184 and VRO42.05.01 from the 
408 MHz and 1420 MHz maps. 
Values given have diffuse background subtracted.
The resulting 408 MHz to 1420 MHz spectral indices, using flux densities without compact sources, 
are 0.25$\pm$0.03 for OA184 and 0.36$\pm$0.06 for VRO42.05.01.  
Table 3 lists the flux densities and spectral indices for both SNRs and flux
densities for the
compact sources within each SNR. 
Compact sources contribute about 21$\%$ at 408 MHz and 12$\%$ at 1420 MHz 
to OA184's flux densities, and have a significant effect on the spectral index of OA184. 
The effects are significantly smaller for VRO42.05.01. 
It is noted that the whole SNR spectral indices derived from integrated flux densities is
consistent with the whole SNR spectral indices derived
by the T-T plot method: 0.23$\pm$0.06 for OA184 and 0.36$\pm$0.03 for VRO42.05.01.

\begin{figure*}
\vspace{55mm}
\begin{picture}(0,80)
\put(-20,-20){\includegraphics{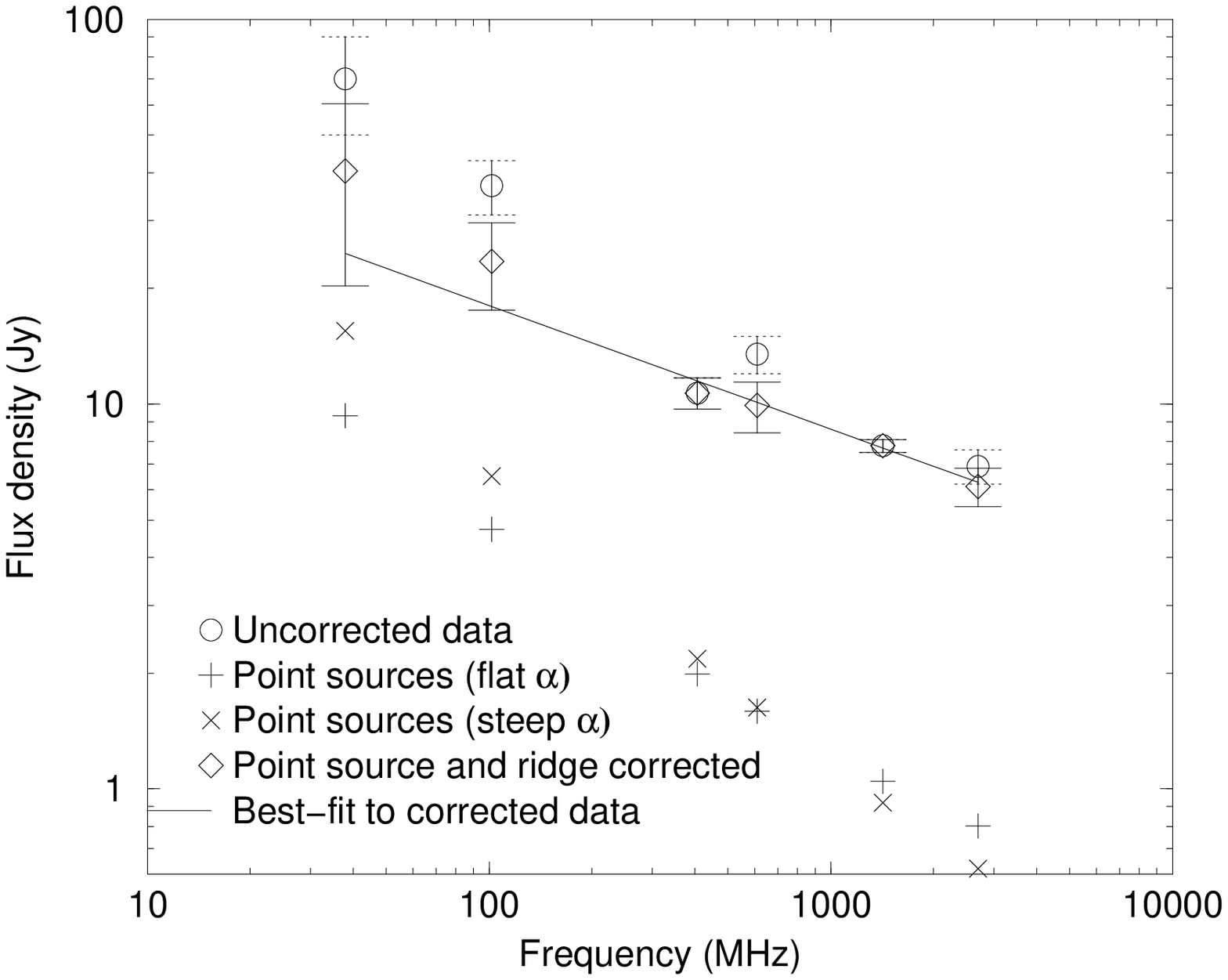}}
\put(240,-20){\includegraphics{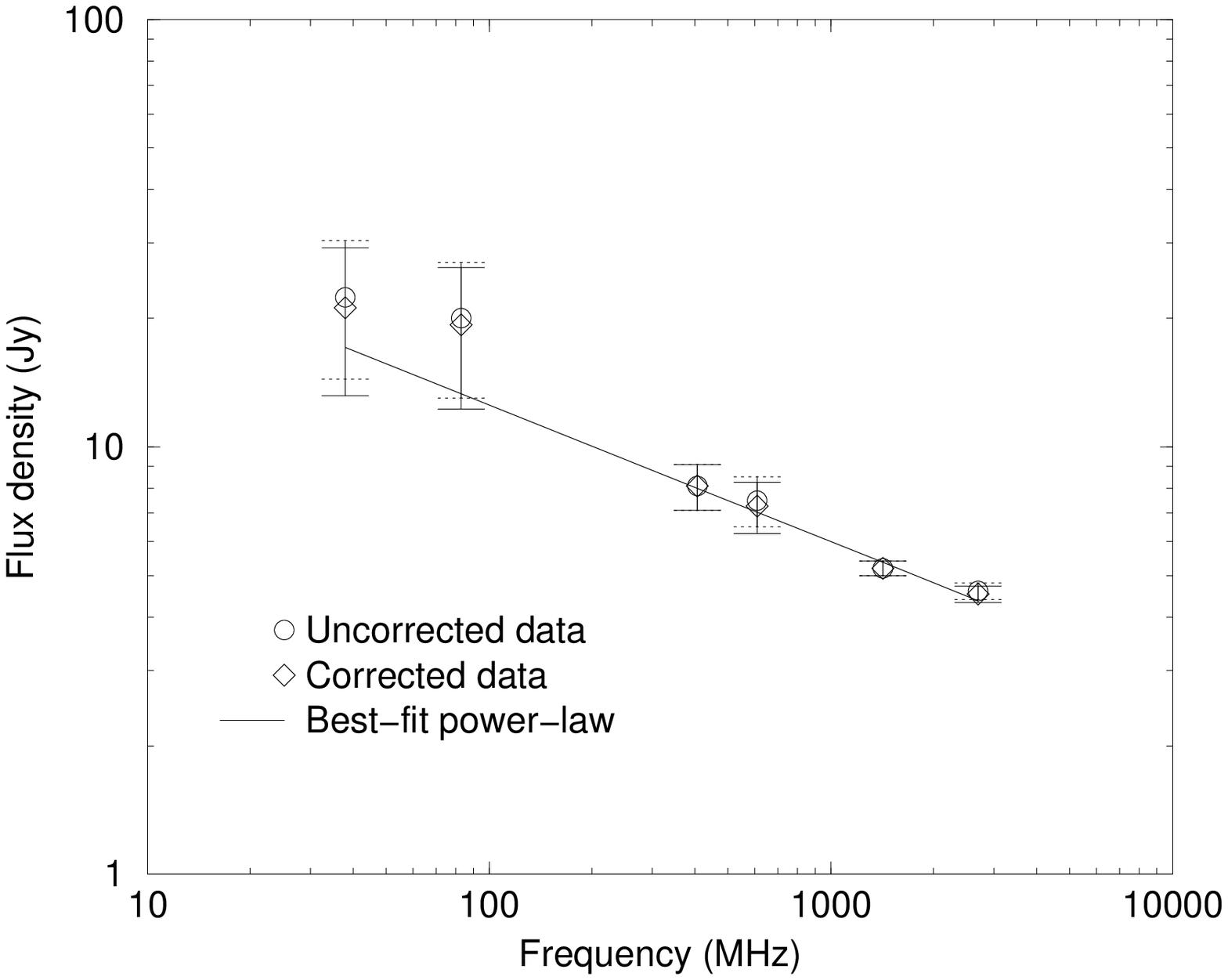}}
\end{picture}
\caption[xx]{Radio spectrum of OA184 (left) and VRO42.05.01 (right). OA184 has a 
best-fit spectral index of 0.32, $\chi^2$ =2.3 for data corrected
for the ridge emission $\alpha_{ridge}$=0.8 and compact sources. 
VRO42.05.01 has best fit spectral index of 0.32, $\chi^2$=2.3 for data corrected 
for compact sources.} 
\end{figure*}

Published integrated flux densities and errors for the two SNRs 
at other frequencies are given in Tables 4 and 5 and shown in Fig. 4. 
We have calculated total compact source flux densities for these other 
frequencies, using the 408-1420 MHz spectral index upper and lower limits and
flux densities from Table 1.
The resulting upper and lower limits to the compact source flux densities within OA184
are shown in Fig. 4 (left). 
For VRO42.04.01 the compact source flux densities are much
smaller, so are not plotted in order to show the SNR flux densities more clearly. 
For OA184, the compact sources' flux density relative to OA184 plus compact sources (plus
ridge when applicable, see below) is the highest at 38 MHz at 
13-22$\%$. It smoothly decreases to 10-13$\%$ at 2695 MHz. For VRO42.04.01, 
it ranges from 5-9$\%$ at 38 MHz to 1.5-2$\%$ at 2695 MHz. 
The published flux densities need to have the compact sources' flux density contribution subtracted
in order to study the spectrum of the SNRs.

For OA184, in addition to the effect of compact sources, there exists the
ridge emission region just west of the SNR, as seen clearly in Fig. 1 at 408 MHz
and 1420 MHz. The ridge is also detected in the Effelsberg 2695 MHz map.
The low resolution published maps include this ridge emission as part of OA184.
For example, the 408 MHz all-sky survey map of Haslam et al. (1982) (with
51$\times$51 arcmin$^{2}$ beam) has a single bright area including both OA184 and 
ridge. 
The 38 MHz, 102 MHz and, to a lesser extent, 610 MHz flux densities need to be 
corrected for the ridge.
The reality of this correction is supported by 4.8 Jy flux density difference between our 
high resolution 408 MHz flux density value and Haslam's low resolution 408 MHz flux 
density value 
(both given in Table 4): 2.2 Jy of the difference is due to compact sources the other
2.6 Jy is due to the ridge (see next paragraph).

We measure the brightness temperatures of the ridge: $\simeq$2.6 K at 408 MHz, 
$\simeq$0.12 K at 1420 MHz, and $\simeq$0.02K in the Effelsberg 2695 MHz map.
The brightness temperatures give the ridge spectral index $\alpha\simeq$0.5 
(408 - 1420 MHz), $\alpha\simeq$0.6 (1420-2695 MHz) and $\alpha\simeq$0.8 
(408 - 2695 MHz). 
These are consistent with eachother within the uncertainties of measurement 
of the ridge brightness temperatures.
The ridge emission region has steeper radio spectrum than the SNR
so it contributes more flux density to OA184 at low frequencies than at high frequencies. 
The ridge flux densities at different frequencies are calculated using the flux 
density at 408 MHz and
spectral index of $\alpha$=0.5-0.8. 
The ridge region flux density is 9-17 Jy at 38 MHz, 5-8 Jy at 102 MHz and 1.9-2.1 Jy at 610 MHz. 

Thus we have recalculated the flux density values of OA184: at 38 MHz, 102 MHz and 610 MHz we 
subtract the ridge flux density and the compact source flux density; 
at 2695 MHz we only subtract the compact
source flux density. 
For 408 and 1420 MHz the values in Table 4 already have compact source flux density
removed and 
 we have already subtracted the ridge flux density by including it in the
background calculation.
For VRO42.05.01, only the compact sources' flux density corrections are needed at 38, 83, 610
and 2700 MHz. We note that Landecker et al. (1982) believe that the flux density at 2700 MHz 
from Willis (1973) is overestimated. Thus we obtain a new flux density value from the
Effelsberg 2695 MHz image: 4.6$\pm$0.2 Jy and use this instead.
We fit the resulting flux density values with a power-law to obtain spectral index. 
Figure 3 shows the corrected flux densities and the best-fit power-law. 
OA184 has a best fit spectral index of $\alpha$=0.32$\pm0.06$, $\chi^2$=2.3, 
when we use a ridge spectral index of 0.8 or $\alpha$=0.34$\pm0.06$, $\chi^2$=4.2, 
when we use a ridge spectral index of 0.5. 
If we omit the most uncertain flux densities of OA184 at the two lowest frequencies, 
the best-fit spectral index is 0.28$\pm0.06$, $\chi^2$=0.4 ( $\alpha_{ridge}$=0.5 or 0.8).  
VRO42.05.01 has a best-fit spectral index of 0.32$\pm0.05$, $\chi^2$=2.3. 

\section{Discussion}

\subsection{OA184}

Low spatial resolution observations of OA184 included the flux density from the
ridge west of OA184 as part of OA184. 
The ridge spectral index of 0.5 to 0.8 is derived from the flux densities in our 
408 and 1420 MHz maps and in the 2695 Effelsberg map. 
The ridge contributes $\simeq$1/3 of the flux density from OA184 in the low frequency maps.
OA184 also has a significant contribution from compact sources within its boundary.
At 408 MHz it is 21$\%$ and at 1420 MHz it is 12$\%$ (Table 3). 
At lower frequencies the contribution is higher: e.g at 38 MHz it is 
$\simeq$17$\%$ of the total or $\simeq$30$\%$ of the OA184 flux density. 
Thus both compact source and ridge emission corrections are necessary to study
the radio spectrum of OA184. 
The resulting multi-frequency spectral index is 0.32$\pm0.06$. This is consistent 
with the 408 MHz - 1420 MHz spectral index derived using an entire SNR
T-T plot method ($\alpha=0.23\pm0.06$) or using integrated flux densities 
($\alpha=0.25\pm0.03$).

OA184 is split into smaller regions: A (east), B (northwest rim), and C
(southwest rim, see Fig. 1) to look for spectral index variations.
The northwest rim (B in Table 2) is found to have a marginally smaller
spectral index (by 1.5 $\sigma$) than whole SNR spectral index. 
Since OA184 is an old remnant (Leahy and Marshall 1988), the shock has slowed
down to a mach number of only a few. Where the shock collides with 
a cooler denser ISM, the mach number will increase resulting in a flatter electron
spectrum and spectral index, as well as a brightening. 
Both are observed at the northwest rim. However, as discussed in 
Leahy and Roger (1998), the effect of mach number on spectral index is only
effective for $\alpha>0.5$ for linear shock acceleration theory.
Non-linear effects in shock acceleration are poorly understood but can reduce spectral index. 
The variable spectral index with $\alpha<0.5$ could
be due to non-linear effects, absorption or an admixture
of thermal radiation which is flattening the spectrum. 
The pre-shock density of OA184 was derived by Leahy and Marshall (1988) 
as 0.2 cm$^{-3}$. This low value and the lack of a low-frequency turnover in
the radio spectrum rule out absorption effects, which have a strong
effect on the low-frequency radio spectrum (Leahy and Roger 1998).  

\subsection{VRO42.05.01}

Compact sources only contribute a few percent to the flux density of VRO42.05.01 
at all frequencies.
The mean spectral index of VRO42.05.01 is 0.36 (Table 3). Yet there are spectral
index variations within this SNR detected at high significance. 
Evidence for this is already seen in 
Fig. 3, lower panels, where several lines of points of different slopes are seen.
For the five subregions (D through H, see Fig. 1), the T-T plots have single
lines of points with well determined slopes (except region E), 
as given in Table 2.
The northeast shell, region D, has the lowest value of $\alpha$, 0.31, and the  
southwest wing, regions F, G and H, have high values, 0.45-0.53.
The boundary region between the shell and the wing, region E, has a T-T plot
consistent with a mixture of $\alpha$=0.31 and 0.47 values. 

This spatial variation of spectral index is consistent with the flux density vs. 
frequency plot. The flux density values (Fig. 4, right) are adequately fit by a single
power law with $\alpha$=0.32, yet it is fit slightly better by a sum of 
two power laws with spectral indices taken from the subregions: $\alpha$=0.31 
(from D) and $\alpha$=0.47 (from F, G and H). 
The possible causes for variable $\alpha<0.5$ are the same as listed
for OA184. For VRO42.05.01, the pre-explosion density is best determined
by X-ray observations and is low- 0.01 to 0.002 cm$^{-3}$ (Burrows
and Guo 1994). This and lack of low-frequency 
turnover in the radio spectrum rule out absorption mechanisms. Also,
the fact that the shell and wing regions have different mach numbers and
different values of $\alpha$ is very suggestive that the spectral index changes 
are due to difference in mach number. To get the spectral indices below 0.5
requires also non-linear shock effects. Further theoretical studies on shock 
acceleration are worthwhile.

\section{Conclusion}

We present new images of the SNRs OA184 and 
VRO42.05.01 at 408 MHz and 1420 MHz from the CGPS data sets
The SNRs' flux densities at both 408 MHz and 1420 MHz are found and
corrected for flux density from compact sources within the SNRs. 
The resulting integrated flux density based spectral index  
is 0.36$\pm$0.06 for VRO42.05.01. 
This agrees with the T-T plot spectral index of 0.36$\pm$0.03
and with the 38 MHz to 2695 MHz spectral index of 0.32$\pm0.05$
from published flux densities corrected for compact sources.
We find that the shell region of VRO42.05.01 has $\alpha$=0.3 
and that the wing region has $\alpha=0.47$. 

For OA184, the integrated flux density based spectral index, 
0.25$\pm$0.03, agrees with the T-T plot spectral index, 0.23$\pm$0.06. 
However, these are smaller than previously published values. 
A new ridge emission region is detected just west of OA184, which was 
included in lower resolution maps as part of OA184. 
Subtracting the ridge and compact source contributions 
yields a spectral index of 0.32$\pm0.06$ (for flux densities from 38 MHz to 2695 MHz) 
or 0.28$\pm0.06$ (for flux densities from 408 MHz to 2695 MHz).
These are now consistent with our 408-1420 MHz spectral indices.
Spatial variations of spectral index are found in OA184: 
$\sim0.1$ for the bright northwest edge and $\sim0.3$ for the rest of the SNR,
although the 
large uncertainties make an accurate determination of this difference 
difficult.
   
\begin{acknowledgements}
We acknowledge support from the Natural Sciences and Engineering Research Council of Canada. W.W. Tian thanks the National Natural Science Foundation of China for support.
The DRAO is operated as a national facility by the National Research Council of Canada. 
The Canadian Galactic Plane Survey is a Canadian project with international partners. 
\end{acknowledgements}

\end{document}